\documentclass[american,aps,prl,twocolumn]{revtex4-1}
\usepackage[T1]{fontenc}
\usepackage[utf8]{inputenc}
\usepackage{babel}
\usepackage{float}
\usepackage{amsmath}
\usepackage{graphicx}
\usepackage[]
 {hyperref}

\makeatletter
\usepackage{graphicx}
\setcitestyle{numbers}

\makeatother

\begin{document}
\title{From $\langle n\rangle$ to $n$: Separating multi-particle dynamics
by intensity cycling}
\author{Pavel Malý$^{1*}$, František Trojánek$^{1}$, Petr Malý$^{1}$ and
Jérémie Léonard$^{2}$}
\affiliation{$^{1}$Faculty of Mathematics and Physics, Charles University, Prague}
\affiliation{$^{2}$IPCMS, Université de Strasbourg -- CNRS, Strasbourg, France\\{*}pavel.maly@matfyz.cuni.cz}
\begin{abstract}
We address the interpretation of transient spectroscopy signals proportional
to average excitation number in terms of discrete number of excitations.
Building on our recent intensity cycling approach, we derive a general
scheme to isolate multi-particle dynamics from any transient signal
under minimal assumptions. The key result is a general formula expressing
the extracted nonlinear signal terms by multi-excitation propagators.
We demonstrate the approach on exciton transport and annihilation
in organic nanoparticles, and carrier recombination in silicon nanocrystals. 
\end{abstract}
\maketitle

Absorption of light creates a number of excitations in any sample
(excitons, charge carriers, polarons, etc.), the dynamics of which
can be excellently tracked by time-resolved spectroscopy.\citep{brixner_exciton_2017,kuhn_quantum_2011}
There, the signal is typically proportional to the average number
of excitations $\langle n(t)\rangle$. In contrast, the signal is
typically interpreted in terms of single excitations, excitation pairs
and so on, relying on a perturbative treatment of the light-matter
interaction.\citep{mukamel_principles_1995} A central question is,
how can the dynamics of a discrete number of particles be obtained
from the signal proportional to $\langle n(t)\rangle$. 

In the dominating linear regime, the average initial number of excitations
$n_{0}=\langle n(t=0)\rangle$ is proportional to the light intensity
$I$, $n_{0}=AI$ with $A$ a constant. Due to the stochastic nature
of light, the number of initially independent excitations is typically
Poisson-distributed, such that the probability to initially excite
$n$ particles is, 

\begin{equation}
P_{n}(t=0)=\frac{n_{0}^{n}}{n!}e^{-n_{0}}=\frac{\left(AI\right)^{n}}{n!}e^{-AI}.\label{eq:poisson}
\end{equation}
During their lifetime, the excitations, however, interact in a stochastic
way, so that the whole changing distribution $P_{n}(t)$ of particle
numbers has to be followed. The time evolution of the distribution
can be described by a hierarchy of coupled differential equations
for the n-particle excitation probabilities $P_{n}(t)$, such as\citep{barzykin_stochastic_2005,jang_effects_1995} 

\begin{align}
\frac{dP_{n}(t)}{dt} & =-k_{1}nP_{n}-\frac{k_{A}(t)}{2}n\left(n-1\right)P_{n}\nonumber \\
 & +k_{1}\left(n+1\right)P_{n+1}+\frac{k_{A}(t)}{2}\left(n+1\right)nP_{n+1}.\label{eq:P-hierarchy}
\end{align}
In this example, $k_{1}$ is a single-excitation recombination rate,
and $k_{A}(t)$ is a two-excitation recombination rate, for excitonic
systems typically exciton--exciton annihilation.\citep{kuhn_quantum_2011}
A solution for the multi-particle populations has been studied for
over fifty years, usually the hierarchy is truncated at some point
by factorization of the multi-particle probability into lower-number
probabilities, resulting in an approximative set of nonlinear differential
equations.\citep{gosele_diffusion_1975,paillotin_analysis_1979,suna_kinematics_1970,jang_effects_1995}
A common severe approximation factorizes the two-particle correlations
into single excitations $\langle n\left(n-1\right)\rangle=\langle n\rangle^{2}$,
producing a closed equation for the average number of excitations,

\begin{equation}
\frac{d\langle n\rangle}{dt}=-k_{1}\langle n\rangle-\frac{k_{A}(t)}{2}\langle n\rangle^{2}.\label{eq:bulk_eea_eq}
\end{equation}
 Since this continuum limit ignores the discrete number of excitations,
it breaks down in any system with only a few interacting excitations
where the stochastic nature of the particle number plays role, leading
to an artificial inferred time dependence of $k_{A}(t)$.\citep{barzykin_stochastic_2007} 

The question addressed in this work is how can the dynamics of single
excitations $P_{1}(t)$, excitation pairs $P_{2}(t)$ and so on obtained
from the spectroscopic signal proportional to $\langle n(t)\rangle$
without relying on such approximations. To this end, we leverage
the known variation of the initial distribution with excitation intensity
$I$ to disentangle the multi-particle dynamics. Thinking in terms
of photon absorption, creating $p$ excitations requires absorption
of $p$ photons, so that this process scales with $p-$th power of
light intensity, $I^{p}$. A separation of the measured signal $S(t)$
into increasingly nonlinear terms in $I$,

\begin{equation}
S(t|I)=S^{(1)}(t)I+S^{(2)}(t)I^{2}+S^{(3)}(t)I^{3}+...\label{eq:S_expansion}
\end{equation}
thus provides an insight into the changing distribution of photon
number. For transient absorption and related spectroscopy techniques,
we have recently introduced a simple, robust approach that we call
intensity cycling.\citep{maly_separating_2023} This method uses measurement
at $N$ intensities to realize this signal decomposition up to $S^{(N)}(t)$
in a sample-independent way. The approach is very simple and works
for any intensities reasonably covering the nonlinear regime.\citep{krich_separating_2025,bruschi_multi-exciton_2026}
First, the signal at $p=1..N$ intensities $I_{p}=\alpha_{p}I$, measured
at multiples of the base intensity $I$, is measured, generating a
linear set of $N$ equations

\begin{equation}
S(t|\alpha_{p}I)=\sum_{n=1}^{N}S^{(n)}(t)\left(\alpha_{p}\right)^{n}I^{n}
\end{equation}
for the nonlinear signals $S^{(n)}I^{n}$ with coefficients $\left(\alpha_{p}\right)^{n}$.
This set of equations can be solved by inversion of the Vandermonde-type
matrix $\left[\alpha_{p}^{n}\right]=\left(\alpha_{p}\right)^{n}$,
obtaining the nonlinear signals as linear combinations of the measured
datasets

\begin{equation}
S^{(n)}(t)I^{n}=\left[\alpha_{p}^{n}\right]^{-1}S(t|\alpha_{p}I).\label{eq:Sn-cycle}
\end{equation}
While this signal decomposition works for any signal in the perturbative
regime, the interpretation of the nonlinear terms $S^{(n)}(t)$ depends
on the system and measurement technique. In our prior work, we expressed
the nonlinear signals in terms of multi-particle dynamics for transient
absorption spectra of specific systems.\citep{maly_separating_2023}
Here, we derive a general, sample and technique independent connection
of the nonlinear signals $S^{(n)}(t)$ and the one- to $N-$particle
dynamics, under the simple assumption of the signal being proportional
to the average number of excitations in the sample, $S(t)=C\langle n(t)\rangle=C\sum_{n=1}^{\infty}nP_{n}(t)$
with $C$ being a constant. The signal decomposition in terms of intensity
is then related to the decomposition of the average excitation number
into the $n-$excitation probabilities $P_{n}(t)$, whose time evolution
is to be determined. The conditional probability of a system initially
with $n$ excitations at time $t=0$ evolving to $p$ excitations
in time $t$ is described by the propagator $U_{pn}(t)$, $P_{p}(t)=\sum_{n=p}^{\infty}U_{pn}(t)P_{n}(t=0)$.
Here, the sum starts at $n=p$ since we do not consider excitation
multiplication in the process, otherwise one would have to start from
$n=1$. Expressing the time-dependent populations by the multi-particle
propagators, and inserting the initial Poissonian distribution (Eq.
(\ref{eq:poisson})) for $P_{n}(t=0)$, we arrive at the expression
for the signal 

\begin{equation}
S(t)=C\langle n(t)\rangle=C\sum_{n=1}^{\infty}\sum_{k\ge n}^{\infty}nU_{nk}(t)\sum_{i=0}^{\infty}\left(-1\right)^{i}\frac{\left(AI\right)^{k+i}}{k!i!}.\label{eq:S(t)_zwischenstep}
\end{equation}
Clearly, we have two expansions of the signal $S(t)$ in terms of
the intensity, Eq. (\ref{eq:S_expansion}) and Eq. (\ref{eq:S(t)_zwischenstep}),
which can be compared term-by-term. Isolating the $n-$th order nonlinear
term imposes $k+i=n$, eliminates one of the summations and imposes
upper bounds for the other two, leading to our central result, an
expression for the $n-$th order nonlinear signal

\begin{equation}
S^{(n)}(t)I^{n}=C\sum_{p=1}^{n}\sum_{k=p}^{n}pU_{pk}(t)\frac{\left(-1\right)^{n-k}}{k!\left(n-k\right)!}\left(AI\right)^{n}.\label{eq:p-thorder signal}
\end{equation}
Notice, that both the sums over $p$ and $k$ are truncated at $n$,
i.e., only a few-particle excitations (one to $n$ excitations) contribute
to $S^{(n)}(t)$. We emphasize that this formula is completely general,
the only assumptions being the signal proportional to $\langle n(t)\rangle$,
initial Poissonian distribution of the excitation number, and number
of excitations not increasing over time. Under these usually met conditions,
the $n-$th order signal $S^{(n)}(t)$ combines dynamics of one to
$n$ excitations according to Eq. (\ref{eq:p-thorder signal}). Specifically,
the first three nonlinear orders that we will use in this work are
of the form

\begin{align}
S^{(1)}(t)I & =Cn_{0}U_{11}(t)\nonumber \\
S^{(2)}(t)I^{2} & =Cn_{0}^{2}\left[-U_{11}(t)+\frac{1}{2}U_{12}(t)+U_{22}(t)\right]\nonumber \\
S^{(3)}(t)I^{3} & =Cn_{0}^{3}\left[\frac{1}{2}U_{11}(t)-\frac{1}{2}U_{12}(t)\right.\label{eq:propagator_combination}\\
 & \left.+\frac{1}{6}U_{13}(t)-U_{22}(t)+\frac{1}{3}U_{23}(t)+\frac{1}{2}U_{33}(t)\right].\nonumber 
\end{align}
Thus, as expected from perturbation theory (i.e., interaction with
photons), the linear signal $S^{(1)}(t)$ reflects single-excitation
dynamics $U_{11}(t)$, the quadratic $S^{(2)}(t)$ features dynamics
of one and two excitations, and so forth. While the expressions in
Eq. (\ref{eq:p-thorder signal}) and Eq. (\ref{eq:propagator_combination})
are completely general, the form of the multi-excitation propagators
$U_{pk}(t)$ depends on the particular system. We illustrate the approach
on two quite different systems and measurements: photoluminescence
(PL) of dye-loaded polymer organic nanoparticles (ONP) with 40 nm
diameter, and transient absorption (TA) of highly confined silicon
nanocrystals 4.8 nm in diameter. 

\begin{figure*}[t]
\includegraphics{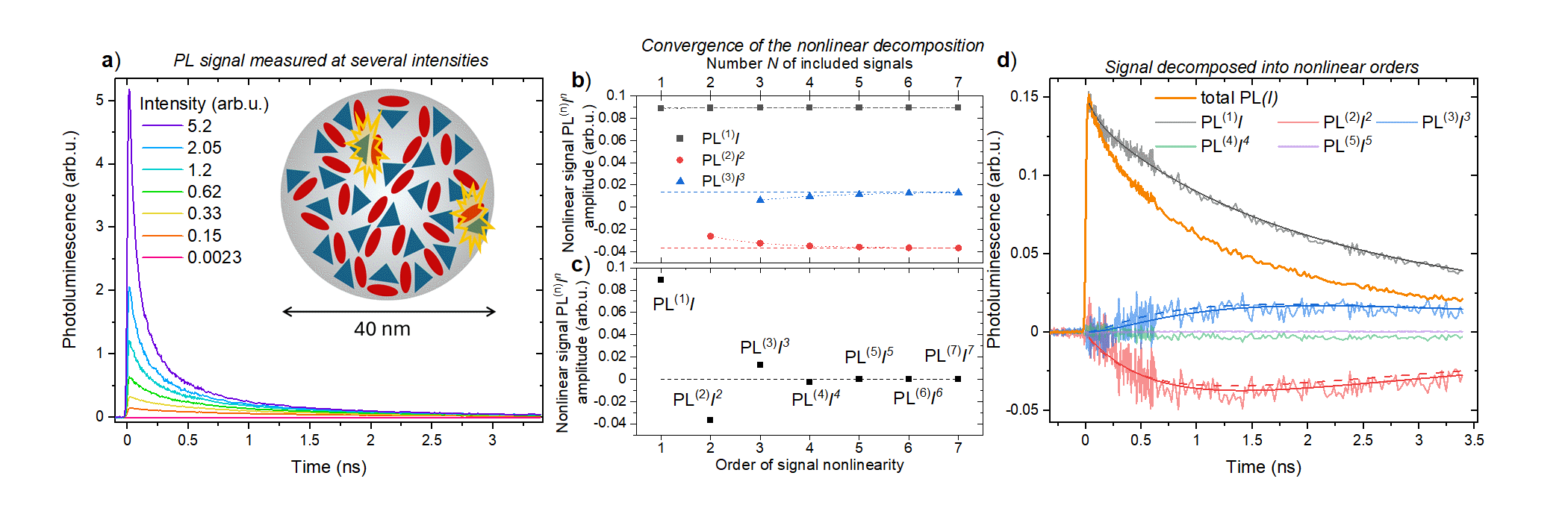}

\caption{Time-resolved photoluminescence (PL) of organic nanoparticles (ONP30)
loaded with rhodamine dye, from our previous work.\citep{gharbi_exciton_2024}
a) Intensity-dependent PL transients, measured by a streak camera
at increasing excitation intensity. Inset: sketch of the ONP with
two excitons. b) Amplitude of the first three nonlinear signals at
1 ns time delay (symbols) extracted by the intensity cycling procedure
at a reference intensity $I=0.15\text{ arb.u.}$. Including progressively
larger $N$ datasets at increasing intensity and thus nonlinear terms,
the nonlinear signals converge to their final value (dashed lines)
as determined by a generic saturation fit (dotted). c) Amplitude of
the nonlinear signals $PL^{(n)}(t=1\text{ ns})I^{n}$, reconstructed
at the reference intensity $I.$ d) Decomposition of the total $PL(t|I)$
signal measured at $I=0.15$ (orange) into the nonlinear orders $PL^{(n)}(t)I^{n}$
(first five orders are shown, terms of higher nonlinearity are negligible,
see Fig. 1c). Smooth curves represent fits of the first three orders
by Eqs. (\ref{eq:n12_kA(t)}),(\ref{eq:n3_kA(t)}), dashed curves
mark the same fit but in addition assuming distribution of ONP sizes.\citep{trofymchuk_giant_2017}
}\label{fig:ONP}
\end{figure*}

The polymer ONPs, sketched in Fig. 1a, are densely loaded by a rhodamine
dye derivative bound to a bulky counter-ion.\citep{trofymchuk_giant_2017}
As a result, the ONPs feature very high exciton diffusivity and efficient
exciton--exciton annihilation (EEA).\citep{gharbi_energy_2023,gharbi_exciton_2024}
 In our previous work, we followed the excitons by intensity-dependent
time-resolved photoluminescence (PL), which is proportional to the
average number of excitons, $PL(t)=C\langle n(t)\rangle$\citep{gharbi_energy_2023}
The PL transients are shown in Fig. 1a, for an increasing intensity
the initial decay becomes faster, reflecting presence of exciton--exciton
annihilation. Using the local character of the annihilation to track
exciton transport has a long tradition,\citep{gosele_diffusion_1975,jang_effects_1995,valkunas_nonlinear_1995}
the physics is captured by the evolution of the spatially-resolved
two-particle density 

\begin{align}
\frac{\partial\rho^{(2)}(t)}{\partial t} & =2D\nabla_{r}^{2}\rho^{(2)}(r,t)-2k_{1}\rho^{(2)}(r,t)\nonumber \\
 & -k_{EEA}(r)\rho^{(2)}(r,t)+f(\rho^{(3)}(r,t)).\label{eq:rho2_diffusive}
\end{align}
This equation captures the interplay of the local annihilation rate
$k_{EEA}(r)$ and exciton transport with diffusion coefficient $D$
. Traditionally, approximations are made for the multi-exciton densities,
such as factorization of the three-particle density $\rho^{(3)}(r,t)$
in Eq. (\ref{eq:rho2_diffusive}) that leads to a closed set of approximative
nonlinear equations. Crucially, in contrast to all preceding works
including our own, thanks to the intensity cycling we probe exclusively
the two-particle dynamics and therefore have access to Eq.\textbf{
}(\ref{eq:rho2_diffusive}) without the three-particle term. The exciton
diffusion coefficient can be inferred from the overall annihilation
rate $k_{A}(t)=\int d^{3}rk_{A}(r)\rho^{(2)}(r,t)$ (one needs to
be careful about normalization of $\rho^{(2)}(r,t)$ here.\citep{suna_kinematics_1970,jang_effects_1995})
 Assuming F\"orster type annihilation rate $k_{EEA}(r)\propto\frac{1}{r^{6}}$,
Eq. (\ref{eq:rho2_diffusive}) can be approximately solved, getting
the time-dependent annihilation rate \citep{gosele_diffusion_1975,jang_effects_1995}

\begin{equation}
k_{A}(t)=4\pi2Dr^{*}\left[1+1.14\frac{r^{*}}{\sqrt{\pi2Dt}}\right].\label{eq:Goesele_kA}
\end{equation}
Here, $r^{*}$ is an effective initial particle interaction radius,
the transient term represents initial fast annihilation of neighboring
excitations, while the constant term reflects the longer-time quasi-stationary
interplay of diffusion and annihilation. 

The crucial advantage of our approach over the previous work \citep{gharbi_exciton_2024}
is that we can infer $k_{A}(t)$ directly from the measured signal.
Instead of the nonlinear approximative equation (\ref{eq:bulk_eea_eq}),
we obtain a finite set of linear differential equations, Eq. (\ref{eq:P-hierarchy}).
This is of the form $\dot{\mathbf{P}}(t)={\cal K}(t)\mathbf{P}(t)$
with a lower-triangular rate matrix ${\cal K}(t)$ and can be analytically
solved by integration and exponentiation of the rate matrix, obtaining
the n-particle propagators $U_{np}(t)$ connecting $p$ excitons at
$t=0$ with $n$ excitons at time $t$. Combining these according
to Eq. (\ref{eq:propagator_combination}), we get for the first three
nonlinear signals

\begin{align}
PL^{(1)}(t)I & =Cn_{0}e^{-k_{1}t}\nonumber \\
PL^{(2)}(t)I^{2} & =-C\frac{n_{0}^{2}}{2}\frac{\int_{0}^{t}k_{A}(\tau)d\tau}{\int_{0}^{t}\left(k_{A}(\tau)+k_{1}\right)d\tau}\nonumber \\
 & \left(1-e^{-\int_{0}^{t}\left(k_{A}(\tau)+k_{1}\right)d\tau}\right)e^{-k_{1}t}\label{eq:n12_kA(t)}
\end{align}

\begin{widetext}

\begin{equation}
PL^{(3)}(t)I^{3}=Cn_{0}^{3}\frac{e^{-3\left(\kappa(t)+\lambda(t)\right)}\lambda(t)^{2}\left[\kappa(t)+\lambda(t)+e^{2\kappa(t)+3\lambda(t)}\left(\kappa(t)+2\lambda(t)\right)-e^{\kappa(t)+2\lambda(t)}\left(2\kappa(t)+3\lambda(t)\right)\right]}{2\left(\kappa(t)+\lambda(t)\right)\left(\kappa(t)+2\lambda(t)\right)\left(2\kappa(t)+3\lambda(t)\right)},\label{eq:n3_kA(t)}
\end{equation}

\end{widetext}where $\lambda(t)=\int_{0}^{t}k_{A}(\tau)d\tau$ and
$\kappa(t)=kt$. A general feature is the alternating sign of the
nonlinear signals, and their scaling with increasing power of $n_{0}$.
The linear signal $PL^{(1)}(t)$ decays with the single-exciton lifetime,
while the quadratic term $PL^{(2)}(t)$ rises from zero with the EEA,
which is the source of nonlinearity. For the common case of EEA faster
than single-excitation lifetime, we have $\int_{0}^{t}\left(k_{A}(\tau)+k_{1}\right)d\tau\approx\int_{0}^{t}k_{A}(\tau)d\tau$,
and the expression approaches the more intuitive form $PL_{\text{fast EEA}}^{(2)}(t)=-C\frac{n_{0}^{2}}{2}\left(1-e^{-\int_{0}^{t}k_{A}(\tau)d\tau}\right)e^{-k_{1}t}$,
which is the expression we have derived in Ref. \citep{maly_separating_2023}
and was since then used by other groups.\citep{shi_annihilation-limited_2025,zhang_probing_2025} 

Now, we extract the nonlinear signals $PL^{(n)}(t)I^{n}$ from the
data in Fig. 1a. We take the second lowest intensity marked by an
estimated excitation density $0.15\cdot10^{-3}\text{ nm}^{-3}$\citep{gharbi_exciton_2024}
as the reference value $I$ and use increasing number $N$ of measured
signals (corresponding to number of terms in the nonlinear expansion
(\ref{eq:S_expansion})) to reconstruct the nonlinear signals $PL^{(n)}(t)I^{n}$
by intensity cycling via Eq. (\ref{eq:Sn-cycle}). In Fig. 1b, we
show the amplitude of the first three nonlinear signals (at delay
of 1 ns), with increasing $N$ the signals converge to their final
values (determined by a phenomenological saturation fit for illustration).
For the linear signal $PL^{(1)}I$ already the lowest-power measurement
provides a good approximation, and from $N=2$ intensities its value
does not change any more. Starting from $N=2$ the quadratic signal
$PL^{(2)}$can be isolated and $N\geq5$ nonlinear terms are needed
for $PL^{(2)}I^{2}$ to converge with the used intensities. The cubic
$PL^{(3)}$ signal needs all $N=7$ intensities, which will thus be
our choice for analysis. For $N=7$, we plot in Fig. 1c the amplitude
of the $n=1..7$ nonlinear signals $PL^{(n)}(t=1\text{ ns})I^{n}$
(at the reference intensity), decreasing in magnitude with the order
of nonlinearity. Note, that while Fig. 1c reports on the convergence
of the nonlinear expansion of the total signal $PL(t)=PL^{(1)}(t)I+PL^{(2)}(t)I^{2}+...$,
Fig. 1b marks the convergence of the individual terms of the expansion.
Finally, in Fig. 1d the total PL signal at the reference intensity
is shown decomposed into the first five nonlinear signals. While immediately
after excitation the linear signal dominates the PL, the nonlinear
signals rise with the exciton interaction, as expected from the theory
in Eqs. (\ref{eq:n12_kA(t)}),(\ref{eq:n3_kA(t)}), which we now use
to fit the data. 

Beginning with the linear signal, we find that already this decays
multi-exponentially, for compatibility with our previous work we fixed
the time constants to 100 ps (0.1 amplitude), 1.35 ns (0.4 amplitude)
and 4.34 ns (0.5 amplitude). The presence of multiple decay rates
in the single-exciton linear signal indicates the presence of a small
number of quenchers.\citep{gharbi_energy_2023,gharbi_exciton_2024}
Assuming three fractions of the ONPs with the respective decay rates
$k_{1}$, we continue with the fitting using Eq. (\ref{eq:Goesele_kA})
for the EEA rate $k_{A}(t)$. We could have taken $r^{*}$ as a fit
parameter, but since we already have a good estimate for the annihilation
radius $R_{EEA}=5.7\text{ nm}$ from our previous work, we use the
well-known approximation \citep{jang_effects_1995}

\begin{equation}
r^{*}=R_{EEA}\frac{\Gamma\left(\frac{3}{4}\right)}{2\Gamma\left(\frac{5}{4}\right)}\left(\frac{k_{1}R_{EEA}^{2}}{2D}\right)^{\frac{1}{4}},\label{eq:rstar}
\end{equation}
where $k_{1}$ is the single-exciton inverse lifetime with the absence
of the quencher and the diffusion coefficient $D$ is still to be
determined. Having fixed $k_{1}$ and $Cn_{0}$ from the linear $PL^{(1)}(t)$
signal, our fit (see Fig. 1d) has only two parameters: $n_{0}$ sensitive
to the nonlinear order scaling and $D$ sensitive to their temporal
evolution. The fit gives the average exciton number per ONP $n_{0}=2.2\pm0.1$.
A clear advantage is that we can determine $n_{0}$ without input
from any other measurements of the beam size, dye absorption cross
section, concentration etc. Our previous estimate of the exciton density,
relying on all these quantities, was $\frac{n_{0}}{V_{ONP}}=0.15\cdot10^{-3}\text{ nm}^{-3}$,
which gives $n_{0}\approx5$, about two times larger than the present
value. The diffusion coefficient comes out $D=674\pm71\text{ nm}^{2}\text{ns}^{-1}$,
which corrects our previous approximative value of $D=370\text{ nm}^{2}\text{ns}^{-1}$.\citep{gharbi_exciton_2024}
Interestingly, the present value is in good agreement with the diffusion
coefficient inferred by an independent PL upconversion anisotropy
measurement.\citep{gharbi_energy_2023} An additional source of uncertainty
is the relatively broad distribution of the ONP sizes.\citep{trofymchuk_giant_2017}.
We tried fitting with such size distribution (see Fig. 1d), obtaining
indeed somewhat larger $n_{0}=2.7\pm0.1$, and a smaller diffusion
coefficient $D=239\pm29\text{ nm}^{2}\text{ns}^{-1}$, close to our
previous results. As a physical consistency check, we evaluate the
value of $r^{*}$ from Eq. (\ref{eq:rstar}), getting $r^{*}=1.1\text{ nm}$
which is very close to the average inter-dye distance of $1.2\text{ nm}.$\citep{gharbi_exciton_2024}
We also note that the time-dependent term in Eq. (\ref{eq:Goesele_kA})
is very small (2\% of the constant term at $t=1\text{ ns}$), and
fitting without it yields nearly identical results, confirming that
the annihilation is diffusion-limited. 

To further demonstrate the range of applicability of the method, we
examined carrier dynamics in silicon nanocrystals (Si-NC), a sample
in which interaction of the highly confined carriers gives rise to
a strong nonlinearity defining the optical response. The Si-NC were
probed by transient absorption pump--probe spectroscopy, in configuration
where the signal is proportional to the average population $\langle n(t)\rangle$
of excited carriers. The analysis of the data followed the same approach
as for the ONPs, with Eq. (\ref{eq:P-hierarchy}) adapted for presence
of three-particle Auger recombination, and is described in the Appendix.
We found that the Si-NC measurement lay on the edge of the intensity
cycling applicability, due to the strong nonlinearity and low transient
signal to noise ratio. While the approach succeeded in separating
the linear carrier trapping from nonlinear recombination, this measurement
serves as an example of a situation in which the intensity cycling
approach begins to fail in reliably separating the multi-particle
signals. 

Nonlinear time-resolved spectroscopy is an ideal tool to probe multi-excitation
dynamics, but its stochastic light excites a distribution of interacting
excitations in the sample. This complicates the interpretation of
the data in terms of one-, two- etc. excitation dynamics. In this
work, we have shown that the recently developed intensity cycling
approach solves this problem, cleanly separating multi-particle dynamics.
The approach is completely general, independent of the probed system,
with the only assumptions being a known initial distribution, recombinative
type of interaction and signal proportional to the number of excitations.
We have demonstrated the approach on two examples: exciton annihilation
and transport in organic nanoparticlers measured by time-resolved
photoluminescence, and carrier recombination in silicon nanocrystals
measured by transient absorption. The approach allowed us to determine
the average number of excitations per system, quantify the single-
and two-particle recombination rates, and infer the diffusion coefficient,
all without input from other measurements. We are convinced that,
due to its robustness, simplicity and universality, the approach will
find application in wide class of spectroscopy techniques of wide
range of systems. 

\setcitestyle{numbers}

\subsection*{Acknowledgements}
\begin{acknowledgments}
PM acknowledges funding by Charles University (grant no. PRIMUS/24/SCI/007),
and by Czech Science Foundation (GA\v{C}R, grant no. 26-23570S). 
\end{acknowledgments}

\subsection*{Data availability}

The data supporting the findings of this study have been in part published
previously (for the ONPs, Ref. \citep{gharbi_exciton_2024}), and
in part (for the Si-NC) are openly available in Zenodo repository
at \href{https://doi.org/10.5281/zenodo.23181121}{https://doi.org/10.5281/zenodo.23181121}.

\subsection*{Appendix}

On the example of exciton transport and annihilation in the ONPs in
the main text, we have demonstrated the utility of the intensity cycling
approach in tracking excitation interaction. Now, we turn to an example
of extremely confined system, carrier dynamics in silicon nanocrystals
probed by transient absorption pump--probe spectroscopy. Here, the
weak signal and strong nonlinearity pushes the approach to its limits.
There is a discussion of the possible modes of carrier recombination
in these silicon nanocrystals.\citep{chlouba_interplay_2018} Following
previous works, we consider three processes: linear recombination
with rate $k_{1}$, two-particle annihilation with rate $\gamma$,
and three-particle Auger recombination with rate $\alpha$. Since
the NCs are small, we neglect carrier transport and take all rates
to be constant.

\begin{figure*}[t]
\includegraphics{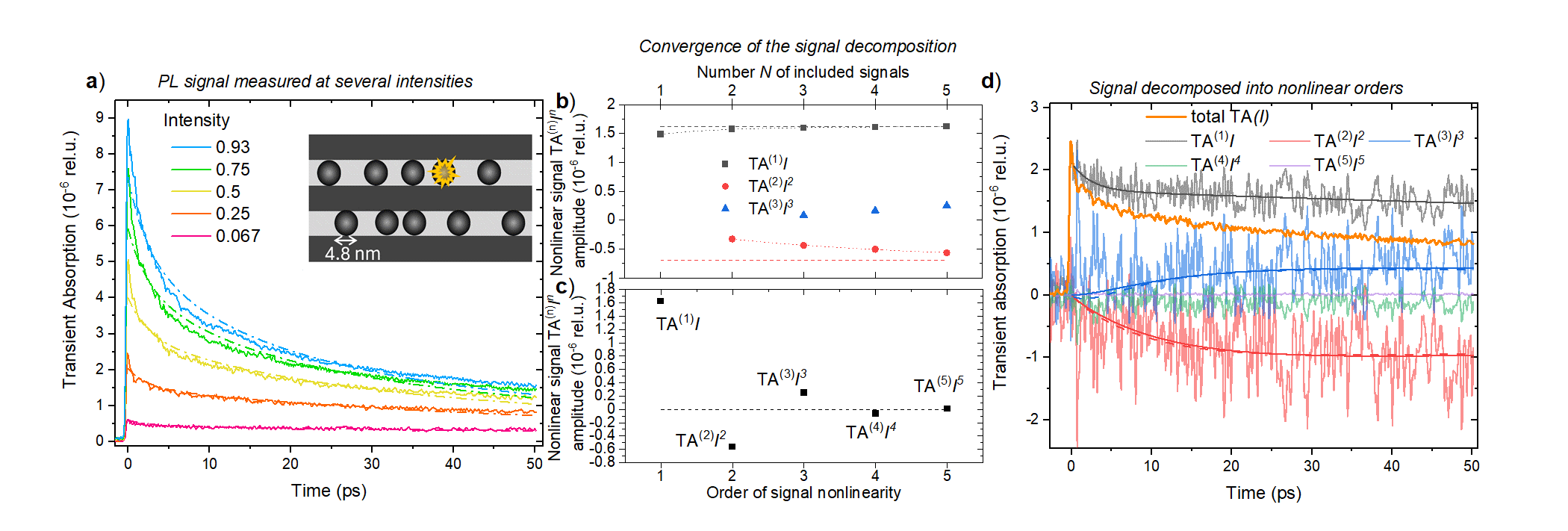}

\caption{Transient absorption (TA) of silicon nanocrystals. a) Intensity-dependent
TA transients, measured with increasing pump excitation intensity.
Inset: sketch of the Si NCs in a layered $\text{SiO}_{2}$ superlattice
structure. Dashed lines: attempted fit by an approximative Eq. (\ref{eq:bulk_eea_eq}).
b) Amplitude of the first three nonlinear signals at 10 ps time delay
(symbols) extracted by the intensity cycling procedure at a reference
intensity $I=0.25\text{ rel.u.}$, when including progressively larger
number $N$ of datasets and thus nonlinear terms. In contrast to Fig.
1b, only the linear term reaches its final value (dashed lines) as
determined by a generic saturation fit (dotted). c) Amplitude of the
nonlinear signals $TA^{(n)}(t=10\text{ ps})I^{n},$ reconstructed
at the reference intensity $I$. d) Decomposition of the total $TA(t|I)$
signal measured at $I=0.25\text{ rel.u.}$ (orange), decomposed into
the nonlinear orders $TA^{(n)}(t)I^{n}$. Smooth curves represent
fits of the first three orders by Eq. (\ref{eq:n2_sol_kA-1}) ($\gamma=0.09\pm0.03\text{ ps}^{-1}$,
and $\alpha=0.0\pm0.02\text{ ps}^{-1}$), dashed curves fits by Eq.
(\ref{eq:n2_sol_kA-1}) with fixed $\gamma=0\text{ ps}^{-1}$ and
$\alpha=0.05\pm0.01\text{ ps}^{-1}$.}\label{fig:SiNC}
\end{figure*}

For $n$ particles, the linear recombination is $nk_{1}$, the two-particle
${n \choose 2}\gamma$, and the three-particle $n{n \choose 2}\alpha$
(reflecting two electrons and one hole, or vice versa). Adding the
three-particle Auger into Eq. (\ref{eq:P-hierarchy}) leads to 

\begin{align}
\frac{dP_{n}(t)}{dt}=-k_{1}nP_{n} & -\frac{\gamma}{2}n\left(n-1\right)P_{n}-\frac{\alpha}{2}n^{2}\left(n-1\right)P_{n}\nonumber \\
+k_{1}\left(n+1\right)P_{n+1} & +\frac{\gamma}{2}\left(n+1\right)nP_{n+1}+\frac{\alpha}{2}\left(n+1\right)^{2}nP_{n+1}.\label{eq:P-hierarchy-Si}
\end{align}
Normally, the analytic solution of this set of equations features
a set of Jacobi polynomials and exponential decays\citep{barzykin_stochastic_2007}.
However, when we separate the nonlinear orders of the signal, we can
truncate the equations by the maximal nonlinear order $N$. Solving
Eq. (\ref{eq:P-hierarchy-Si}) for $n=1,2,3$ and plugging the propagators
into Eq. (\ref{eq:propagator_combination}), we obtain the nonlinear
TA signal terms
\begin{align}
TA^{(1)}(t)I & =Cn_{0}e^{-k_{1}t}\nonumber \\
TA^{(2)}(t)I^{2} & =-Cn_{0}^{2}\frac{\gamma+2\alpha}{2\left(k_{1}+\gamma+2\alpha\right)}\left(1-e^{-\left(\gamma+2\alpha+k_{1}\right)t}\right)e^{-k_{1}t},\label{eq:n2_sol_kA-1}
\end{align}

\begin{widetext}

\begin{align}
TA^{(3)}(t)I^{3} & =Cn_{0}^{3}\frac{1}{2}\left[\frac{\left(15\alpha^{2}+\alpha(8\gamma+k_{1})+\gamma^{2}\right)e^{-3t(3\alpha+\gamma+\text{\ensuremath{k_{1}}})}}{(7\alpha+2\gamma+k_{1})(9\alpha+3\gamma+2\text{\ensuremath{k_{1}}})}+\frac{e^{-\text{\ensuremath{k_{1}}}t}\left(24\alpha^{2}+\alpha(20\gamma+21\text{\ensuremath{k_{1}}})+4(\gamma+\text{\ensuremath{k_{1}}})^{2}\right)}{(2\alpha+\gamma+\text{\ensuremath{k_{1}}})(9\alpha+3\gamma+2\text{\ensuremath{k_{1}}})}\right.\label{eq:TA3}\\
 & -\frac{\left(34\alpha^{2}+\alpha(27\gamma+26\text{\ensuremath{k_{1}}})+(\gamma+\text{\ensuremath{k_{1}}})(5\gamma+4\text{\ensuremath{k_{1}}})\right)e^{-(t(2\alpha+\gamma+2k_{1}))}}{(2\alpha+\gamma+\text{\ensuremath{k_{1}}})(7\alpha+2\gamma+\text{\ensuremath{k_{1}}})}+\frac{2(3\alpha+\gamma+\text{\ensuremath{k_{1}}})e^{-(t(2\alpha+\gamma+2\text{\ensuremath{k_{1}}}))}}{7\alpha+2\gamma+\text{\ensuremath{k_{1}}}}\nonumber \\
 & \left.+\frac{(2\alpha+\gamma+2\text{\ensuremath{k_{1}}})e^{-(t(2\alpha+\gamma+2\text{\ensuremath{k_{1}}}))}}{2\alpha+\gamma+\text{\ensuremath{k_{1}}}}-\frac{e^{-\text{\ensuremath{k_{1}}}t}(2\alpha+\gamma+2\text{\ensuremath{k_{1}}})}{2\alpha+\gamma+\text{\ensuremath{k_{1}}}}\right].
\end{align}

\end{widetext}

As expected, the linear signal decays with the single-particle lifetime
only. In the quadratic signal, we find both the two- and three-particle
decays, because particles of two types are considered - two photons
are sufficient to produce two electron-hole pairs, providing the three
interacting particles. If the interacting term was $\propto{n \choose 3}$,
such recombination would first appear in the cubic signal. 

The TA on the Si NCs was measured with a standard pump--probe setup,
using Ti:Sapphire oscillator and amplifier (Spectra Physics) providing
pulses at 1 kHz repetition rate.\citep{chlouba_interplay_2018} A
second harmonic at 400 nm was used for pumping above the bandgap,
and the fundamental at 800 nm for probing by sub-bandgap light. In
this configuration, the signal is proportional to the average number
of excited electron-hole pairs, $TA(t)=C\langle n\rangle.$ The transient
signal was detected by a photodiode and isolated by a lock-in detector
at the frequency of the chopped pump beam. The measured transients
are shown in Fig. 2a, along with a sketch of the nanocrystal sample
comprising silicon nanocrystals 4.8 nm in diameter within 40 $\text{SiO}_{0.93}$/$\text{SiO}_{2}$layers. 

We performed the same intensity cycling analysis as before for the
ONPs, choosing the intensity of $I=0.25\text{ rel.u.}$ as the reference
value. This time, the amplitude of the nonlinear terms around 10 ps,
shown in Fig. 2b, converges more slowly with the number of considered
intensities, basically only the linear term is isolated faithfully,
the quadratic signal reached only 80\% of its expected value (based
on a simple saturation fit), and for the cubic signal no convergence
could be verified, indicating that the magnitude of the $TA^{(3)}(t)I^{3}$
signal is underestimated. Still, taking the $N=5$ datasets, the nonlinear
signals do decrease in magnitude sufficiently (Fig. 2c) for the decomposition
of the total signal at the reference intensity to be feasible. This
decomposition of the total $TA(t)$ signal is shown in Fig. 2d, as
before, the initially dominating linear signal is quickly accompanied
by rising nonlinear signals due to the particle interaction. 

We fit the extracted nonlinear orders by our analytical solution,
Eqs. (\ref{eq:n2_sol_kA-1}),(\ref{eq:TA3}), starting with the the
linear signal. The $TA^{(1)}(t)$ exhibited, perhaps surprisingly,
a two-exponential decay, 24\% of the signal decayed with time constant
of $2.4\pm0.7\text{ ps}$, while the remaining 76\% decayed much more
slowly with a decay time of $384\pm78\text{ ps}$ (difficult to determine
in our 50 ps measurement window). In line with previous literature,
we explain the 2.4 ps decay by presence of traps in fraction the nanocrystals.\citep{de_jong_trapping_2017}.
Further fits were thus done with a sum of two NC populations, with
the two decay rates $k_{1}$. Note, that if we fitted directly the
intensity dependent nonlinear signals with a stochastic model, this
trapping of the single carriers could easily be missed, attributing
the signal decay to nonlinear recombination instead. Despite the significant
noise, the nonlinear signals can be fitted by Eqs. (\ref{eq:n2_sol_kA-1}),(\ref{eq:TA3}),
getting $n_{0}=1.37\pm0.04$ excitations per nanocrystal, and the
recombination rates $\gamma=0.09\pm0.03\text{ ps}^{-1}$, and $\alpha=0.0\pm0.02\text{ ps}^{-1}$.
There is, however, a significant dependence between the $\gamma$
and $\alpha$ parameters, and a reasonable fit can be achieved with
fixed $\gamma=0\text{ ps}^{-1}$, yielding $\alpha=0.05\pm0.01$.
While there is a clear difference in the theoretical curves for $\gamma=0$
and $\alpha=0$, differing in their sign at the short times, within
the error margin we cannot reliably distinguish the bi-molecular and
Auger recombination. The recombination constants are often given in
literature per carrier density. The NC volume is $V_{NC}=5.8\cdot10^{-20}\text{cm}^{3}$,
the carrier density at the reference intensity is $n_{0}=2.4\cdot10^{19}\text{cm}^{-3}$
and the bulk recombination rates $\gamma_{B}=\gamma V_{NC}=5.2\cdot10^{-9}\text{ cm}^{3}\text{s}^{-1}$
and $\alpha_{B}=\alpha V_{NC}^{2}\approx1.7\cdot10^{-28}\text{cm}^{6}\text{s}^{-1}$.
All these values are highly comparable to the values of $n_{0}\approx9\cdot10^{19}\text{ cm}^{-3}$,
$\gamma_{B}\approx5\cdot10^{-9}\text{ cm}^{3}\text{s}^{-1}$ and $\alpha_{B}\approx4\cdot10^{-28}\text{cm}^{6}\text{s}^{-1}$
reported by Chlouba et al. for similar NCs under comparable measurement
conditions.\citep{chlouba_interplay_2018}

To demonstrate the necessity of the discrete multi-particle approach,
we tried globally fitting the transients by the volume equation (\ref{eq:bulk_eea_eq})
with constant $k_{A}(t)=\gamma$, fixing the two fractions with the
two linear decay rates $k_{1}$ (see Fig. 2a). As in previous works,\citep{chlouba_interplay_2018}
we find that this description does not fit. However, instead of indicating
the necessity of three-particle interaction, we argue that this points
to the need of the stochastic description by Eq. (\ref{eq:P-hierarchy-Si}),
and the data are compatible with bi-molecular recombination only. 

Rather then yielding accurate recombination rates, the silicon nanocrystals
exemplify a case for which the applicability of the intensity cycling
is on the edge. The challenge is the high nonlinearity already in
a very small transient signal (relative change $\frac{\Delta T}{T}\approx10^{-6}$).
Reliable extraction of the nonlinear signals would need measurement
at even lower intensity, for which the signal drowns in noise completely.
Physically, this means that the measurable signal is dominated by
nonlinear effects, preventing its reliable decomposition. Still, even
in this situation the presence of a picosecond decay in the linear
signal is a useful information. 

\bibliographystyle{apsrev4-1}
\bibliography{MPIcycle}

\end{document}